\documentclass[aps,nofootinbib,preprint,superscriptaddress]{revtex4}%
\usepackage{hyperref}
\usepackage{amsmath}
\usepackage{amsfonts}
\usepackage{amssymb}
\usepackage{graphicx}
\usepackage{color}%
\providecommand{\U}[1]{\protect\rule{.1in}{.1in}}

\begin{document}
\title{The Lauricella Functions and Exact String Scattering Amplitudes}
\author{Sheng-Hong Lai}
\email{xgcj944137@gmail.com}
\affiliation{Department of Electrophysics, National Chiao-Tung University, Hsinchu, Taiwan, R.O.C.}
\author{Jen-Chi Lee}
\email{jcclee@cc.nctu.edu.tw}
\affiliation{Department of Electrophysics, National Chiao-Tung University, Hsinchu, Taiwan, R.O.C.}
\author{Yi Yang}
\email{yiyang@mail.nctu.edu.tw}
\affiliation{Department of Electrophysics, National Chiao-Tung University, Hsinchu, Taiwan, R.O.C.}
\author{}
\date{\today }

\begin{abstract}
We discover that the $26D$ open bosonic string scattering amplitudes (SSA) of
three tachyons and one arbitrary string state can be expressed in terms of the
D-type Lauricella functions with associated $SL(K+3,%
\mathbb{C}
)$ symmetry. As a result, SSA and symmetries or relations among SSA of
different string states at various limits calculated previously can be
rederived. These include the linear relations first conjectured by Gross
\cite{GM,Gross,GrossManes} and later corrected and proved in
\cite{ChanLee1,ChanLee2,CHL,PRL, CHLTY,susy} in the hard scattering limit, the
recurrence relations in the Regge scattering limit with associated $SL(5,%
\mathbb{C}
)$ symmetry \cite{KLY,LY,LY2014} and the extended recurrence relations in the
nonrelativistic scattering limit with associated $SL(4,%
\mathbb{C}
)$ symmetry \cite{LLY} discovered recently. Finally, as an application, we
calculate a new recurrence relation of SSA which is valid for \textit{all} energies.

\end{abstract}
\maketitle

\section{Introduction}

It has long been believed that there exist huge hidden spacetime symmetries of
string theory. As a consistent theory of quantum gravity, string theory
contains no free parameter and an infinite number of higher spin string
states. On the other hand, the very soft exponential fall-off behavior of
string scattering amplitudes (SSA) in the hard scattering limit, in contrast
to the power law behavior of hard field theory scattering amplitudes, strongly
suggests the existence of infinite number of relations among SSA of different
string states. These relations or symmetries soften the UV structure of
quantum string theory. Indeed, this kind of infinite relations were first
conjectured by Gross \cite{GM,Gross,GrossManes} and later corrected and
explicitly proved in \cite{ChanLee1,ChanLee2,CHL,PRL, CHLTY,susy} by using
decoupling of zero-norm states (ZNS) \cite{ZNS1}, and can be used to reduce
the number of independent hard SSA from $\infty$ down to $1$.

It was important to note that the linear relations obtained by decoupling of
ZNS in the hard scattering limit corrected \cite{ChanLee1,ChanLee2,CHL} the
saddle point calculations of Gross \cite{Gross}, Gross and Mende \cite{GM} and
Gross and Manes \cite{GrossManes}. The results of the former authors were
consistent with the decoupling of high energy ZNS or unitarity of the theory
while those of the latter were not. See one simple example to be presented in
Eq.(\ref{03}) in section IV. Independently, the inconsistency of the saddle
point calculations of the above authors was also pointed out by the authors of
\cite{West2} using the group theoretic approach of string amplitudes
\cite{NW5}.

On the other hand, inspired by Witten's seminal paper \cite{witten}, there
have been tremendous developments on calculations of higher point and higher
loop Yang-Mills and gravity field theory amplitudes \cite{JA}. Many
interesting relations among these field theory amplitudes have also been
proposed and suggested. In addition, connections between field theory and
string theory amplitudes are currently under many investigations.

Historically, there were at least three approaches to probe stringy symmetries
or relations among scattering amplitudes of higher spin string states. These
include the gauge symmetry of Witten string field theory, the conjecture of
Gross \cite{Gross} on symmetries or linear relations among SSA of different
string states in the hard scattering limit by the saddle point method
\cite{GM,Gross,GrossManes} and Moore's bracket algebra approach
\cite{Moore,Moore1,CKT} of stringy symmetries. See a recent review
\cite{review} for some connections of these three approaches.

Recently, it was found that the Regge SSA of three tachyons and one arbitrary
string states can be expressed in terms of a sum of Kummer functions $U$
\cite{KLY,LY,LY2014}, which soon later were shown to be the first Appell
function $F_{1}$ \cite{LY2014}. Regge stringy symmetries or recurrence
relations \cite{LY,LY2014} were then constructed and used to reduce the number
of independent\ Regge SSA from $\infty$ down to $1$. Moreover, an interesting
link between Regge SSA and hard SSA was found \cite{KLY,LYY}, and for each
mass level the ratios among hard SSA can be extracted from Regge SSA. This
result enables us to argue that the known $SL(5;C)$ dynamical symmetry of the
Appell function $F_{1}$ \cite{sl5c} is crucial to probe high energy spacetime
symmetry of string theory.

More recently, the extended recurrence relations \cite{LLY} among
nonrelativistic low energy SSA of a class of string states with different
spins and different channels were constructed by using the recurrence
relations of the Gauss hypergeometric functions with associated $SL\left(  4,%
\mathbb{C}
\right)  $ symmetry \cite{sl4c}. These extended recurrence relations
generalize and extend the field theory BCJ \cite{BCJ} relations to higher mass
and higher spin string states.

To further uncover the structure of stringy symmetries, in section II of this
paper we calculate the $26D$ open bosonic SSA of three tachyons and one
arbitrary string states at \textit{arbitrary} energies. We discover that these
SSA can be expressed in terms of the D-type Lauricella functions with
associated $SL(K+3,%
\mathbb{C}
)$ symmetry \cite{sl4c}. As a result, all these SSA and symmetries or
relations among SSA of different string states at various limits calculated
previously can be rederived. These will be presented in sections III, IV and V
which include the recurrence relations in the Regge scattering limit
\cite{LY,LY2014} with associated $SL(5;C)$ symmetry, the linear relations
conjectured by Gross \cite{Gross} and corrected and proved in
\cite{ChanLee1,ChanLee2,CHL,PRL, CHLTY,susy} in the hard scattering limit and
the extended recurrence relations in the nonrelativistic scattering limit
\cite{LLY} with associated $SL(4;C)$ symmetry discovered very recently.
However, since \textit{not} all Lauricella functions $F_{D}^{(K)}$ with
arbitrary \textit{independent} arguments can be used to represent SSA, it
remained to be studied how the basis states of each $SL(K+3,%
\mathbb{C}
)$ group representation for a given $K$ relates to SSA \cite{Kawai}.

As a byproduct from the calculation of rederiving linear relations in the hard
scattering limit directly from Lauricella functions, we propose an identity
Eq.(\ref{pro}) which generalizes the Stirling number identity
Eq.(\ref{stirling}) \cite{KLY,LYY} used previously to extract ratios among
hard SSA from the Appell functions in Regge SSA. Finally, as an example, in
section VI we calculate a new recurrence relation of SSA which is valid for
\textit{all} energies.

\section{Four-point string amplitudes}

We will consider SSA of three tachyons and one arbitrary string states put at
the second vertex. For the 26D open bosonic string, the general states at mass
level $M_{2}^{2}=2(N-1)$, $N=\sum_{n,m,l>0}\left(  nr_{n}^{T}+mr_{m}%
^{P}+lr_{l}^{L}\right)  $ with polarizations on the scattering plane are of
the form%
\begin{equation}
\left\vert r_{n}^{T},r_{m}^{P},r_{l}^{L}\right\rangle =\prod_{n>0}\left(
\alpha_{-n}^{T}\right)  ^{r_{n}^{T}}\prod_{m>0}\left(  \alpha_{-m}^{P}\right)
^{r_{m}^{P}}\prod_{l>0}\left(  \alpha_{-l}^{L}\right)  ^{r_{l}^{L}}%
|0,k\rangle. \label{state}%
\end{equation}
In the CM frame, the kinematics are defined as%
\begin{align}
k_{1}  &  =\left(  \sqrt{M_{1}^{2}+|\vec{k_{1}}|^{2}},-|\vec{k_{1}}|,0\right)
,\\
k_{2}  &  =\left(  \sqrt{M_{2}+|\vec{k_{1}}|^{2}},+|\vec{k_{1}}|,0\right)  ,\\
k_{3}  &  =\left(  -\sqrt{M_{3}^{2}+|\vec{k_{3}}|^{2}},-|\vec{k_{3}}|\cos
\phi,-|\vec{k_{3}}|\sin\phi\right)  ,\\
k_{4}  &  =\left(  -\sqrt{M_{4}^{2}+|\vec{k_{3}}|^{2}},+|\vec{k_{3}}|\cos
\phi,+|\vec{k_{3}}|\sin\phi\right)
\end{align}
with $M_{1}^{2}=M_{3}^{2}=M_{4}^{2}=-2$ and $\phi$ is the scattering angle.
The Mandelstam variables are $s=-\left(  k_{1}+k_{2}\right)  ^{2}$,
$t=-\left(  k_{2}+k_{3}\right)  ^{2}$ and $u=-\left(  k_{1}+k_{3}\right)
^{2}$. There are three polarizations on the scattering plane%
\begin{align}
e^{T}  &  =(0,0,1),\\
e^{L}  &  =\frac{1}{M_{2}}\left(  |\vec{k_{1}}|,\sqrt{M_{2}+|\vec{k_{1}}|^{2}%
},0\right)  ,\\
e^{P}  &  =\frac{1}{M_{2}}\left(  \sqrt{M_{2}+|\vec{k_{1}}|^{2}},|\vec{k_{1}%
}|,0\right)  .
\end{align}
For later use, we define%
\begin{equation}
k_{i}^{X}\equiv e^{X}\cdot k_{i}\text{ \ for \ }X=\left(  T,P,L\right)  .
\end{equation}
Note that SSA of three tachyons and one arbitrary string state with
polarizations orthogonal to the scattering plane vanish.

For illustration, we begin with a simple case, namely, four-point function
with the three tachyons and the highest spin state at mass level $M_{2}%
^{2}=2(N-1)$, $N=p+q+r$ of the following form%
\begin{equation}
\left\vert p,q,r\right\rangle =\left(  \alpha_{-1}^{T}\right)  ^{p}\left(
\alpha_{-1}^{P}\right)  ^{q}\left(  \alpha_{-1}^{L}\right)  ^{r}|0,k\rangle.
\end{equation}
The four-point scattering amplitude can be calculated as%
\begin{align}
A_{st}^{(p,q,r)}  &  =\frac{\sin(\pi k_{2}\cdot k_{4})}{\sin(\pi k_{1}\cdot
k_{2})}A_{tu}^{(p,q,r)}=\frac{\sin(\frac{u}{2}+2-N)\pi}{\sin(\frac{s}%
{2}+2-N)\pi}A_{tu}^{(p,q,r)}\nonumber\\
&  =\frac{(-1)^{N}\Gamma(\frac{s}{2}+2-N)\Gamma(\frac{-s}{2}-1+N)}%
{\Gamma(\frac{u}{2}+2)\Gamma(\frac{-u}{2}-1)}A_{tu}^{(p,q,r)}\nonumber\\
&  =\frac{(-1)^{N}\Gamma(\frac{s}{2}+2-N)\Gamma(\frac{-s}{2}-1+N)}%
{\Gamma(\frac{u}{2}+2)\Gamma(\frac{-u}{2}-1)}\nonumber\\
&  \times\int_{1}^{\infty}dx\,x^{k_{1}\cdot k_{2}}(x-1)^{k_{2}\cdot k_{3}%
}\cdot\left[  \frac{k_{1}^{T}}{x}+\frac{k_{3}^{T}}{x-1}\right]  ^{p}%
\nonumber\\
&  \cdot\left[  \frac{k_{1}^{P}}{x}+\frac{k_{3}^{P}}{x-1}\right]  ^{q}%
\cdot\left[  \frac{k_{1}^{L}}{x}+\frac{k_{3}^{L}}{x-1}\right]  ^{r}\nonumber\\
&  =\frac{\Gamma(\frac{s}{2}+2-N)\Gamma(\frac{-s}{2}-1+N)}{\Gamma(\frac{u}%
{2}+2)\Gamma(\frac{-u}{2}-1)}\left(  -k_{3}^{T}\right)  ^{p}\left(  -k_{3}%
^{P}\right)  ^{q}\left(  -k_{3}^{L}\right)  ^{r}\nonumber\\
&  \times\int_{1}^{\infty}dx\,x^{k_{1}\cdot k_{2}}(x-1)^{k_{2}\cdot k_{3}%
}\cdot\left(  1-(\frac{-k_{1}^{T}}{k_{3}^{T}}))\frac{x-1}{x}\right)
^{p}\nonumber\\
&  \cdot\left(  1-(\frac{-k_{1}^{P}}{k_{3}^{P}})\frac{x-1}{x}\right)
^{q}\cdot\left(  1-(\frac{-k_{1}^{L}}{k_{3}^{L}})\frac{x-1}{x}\right)  ^{r}.
\end{align}
In the above calculation, we have used the string BCJ relation%
\begin{equation}
A_{st}^{(p,q,r)}=\frac{\sin(\pi k_{2}\cdot k_{4})}{\sin(\pi k_{1}\cdot k_{2}%
)}A_{tu}^{(p,q,r)},
\end{equation}
which was proved by monodromy of integration of string amplitudes
\cite{stringBCJ, stringBCJ2} and explicitly proved recently in \cite{LLY}. We
can now do a change of variable $\frac{x-1}{x}=x^{\prime}$ to get%

\begin{align}
A_{st}^{(p,q,r)}  &  =\frac{\Gamma(\frac{s}{2}+2-N)\Gamma(\frac{-s}{2}%
-1+N)}{\Gamma(\frac{u}{2}+2)\Gamma(\frac{-u}{2}-1)}\left(  -k_{3}^{T}\right)
^{p}\left(  -k_{3}^{P}\right)  ^{q}\left(  -k_{3}^{L}\right)  ^{r}\nonumber\\
&  \times\int_{0}^{1}dx^{\prime}\,x^{\prime\frac{-t}{2}-2}(1-x^{\prime
})^{\frac{-u}{2}-2}\left(  1-(\frac{-k_{1}^{T}}{k_{3}^{T}})x^{\prime}\right)
^{p}\nonumber\\
&  \cdot\left(  1-(\frac{-k_{1}^{P}}{k_{3}^{P}})x^{\prime}\right)  ^{q}%
\cdot\left(  1-(\frac{-k_{1}^{L}}{k_{3}^{L}})x^{\prime}\right)  ^{r}%
\nonumber\\
&  =\frac{\Gamma(\frac{s}{2}+2-N)\Gamma(\frac{-s}{2}-1+N)}{\Gamma(\frac{u}%
{2}+2)\Gamma(\frac{-u}{2}-1)}\nonumber\\
&  \cdot\left(  -k_{3}^{T}\right)  ^{p}\left(  -k_{3}^{P}\right)  ^{q}\left(
-k_{3}^{L}\right)  ^{r}\frac{\Gamma(\frac{-t}{2}-1)\Gamma(\frac{-u}{2}%
-1)}{\Gamma(\frac{s}{2}+2-N)}\nonumber\\
&  \times F_{D}^{(3)}(\frac{-t}{2}-1,-p,-q,-r,\frac{s}{2}+2-N;\frac{-k_{1}%
^{T}}{k_{3}^{T}},\frac{-k_{1}^{P}}{k_{3}^{P}},\frac{-k_{1}^{L}}{k_{3}^{L}}),
\end{align}
which can be written as%

\begin{align}
A_{st}^{(p,q,r)}  &  =\left(  -k_{3}^{T}\right)  ^{p}\left(  -k_{3}%
^{P}\right)  ^{q}\left(  -k_{3}^{L}\right)  ^{r}\frac{\Gamma(\frac{-s}%
{2}-1+N)\Gamma(\frac{-t}{2}-1)}{\Gamma(\frac{u}{2}+2)}\nonumber\\
&  \times F_{D}^{(3)}(\frac{-t}{2}-1,-p,-q,-r,\frac{s}{2}+2-N;-C^{T}%
,-C^{P},-C^{L}) \label{111}%
\end{align}
if we define%
\begin{equation}
k_{i}^{X}=e^{X}\cdot k_{i}\text{, \ \ \ \ }\frac{k_{3}^{X}}{k_{1}^{X}}%
=C^{X}\text{.}%
\end{equation}
In Eq.(\ref{111}), the D-type Lauricella function $F_{D}^{(K)}$ is one of the
four extensions of the Gauss hypergeometric function to $K$ variables and is
defined as%
\begin{align}
&  F_{D}^{(K)}\left(  a;b_{1},...,b_{K};c;x_{1},...,x_{K}\right) \nonumber\\
=  &  \sum_{n_{1},\cdots,n_{K}}\frac{\left(  a\right)  _{n_{1}+\cdots+n_{K}}%
}{\left(  c\right)  _{n_{1}+\cdots+n_{K}}}\frac{\left(  b_{1}\right)  _{n_{1}%
}\cdots\left(  b_{K}\right)  _{n_{K}}}{n_{1}!\cdots n_{K}!}x_{1}^{n_{1}}\cdots
x_{K}^{n_{K}}%
\end{align}
where $(a)_{n}=a\cdot\left(  a+1\right)  \cdots\left(  a+n-1\right)  $ is the
Pochhammer symbol. There is a integral representation of the Lauricella
function $F_{D}^{(K)}$ discovered by Appell and Kampe de Feriet (1926)
\cite{Appell}%
\begin{align}
&  F_{D}^{(K)}\left(  a;b_{1},...,b_{K};c;x_{1},...,x_{K}\right) \nonumber\\
&  =\frac{\Gamma(c)}{\Gamma(a)\Gamma(c-a)}\int_{0}^{1}dt\,t^{a-1}%
(1-t)^{c-a-1}\nonumber\\
&  \cdot(1-x_{1}t)^{-b_{1}}(1-x_{2}t)^{-b_{2}}...(1-x_{K}t)^{-b_{K}},
\label{Lau}%
\end{align}
which can be used to directly calculate the amplitude in Eq.(\ref{111}). The
relevance of the Lauricella function in Eq.(\ref{Lau}) for string scattering
amplitudes was first suggested in \cite{LY2014}.

We now calculate the string four-point scattering amplitude with three
tachyons and one general higher spin state in Eq.(\ref{state}) as following%
\begin{align}
A_{st}^{(p_{n};q_{m};r_{l})}  &  =\frac{\sin(\pi k_{2}\cdot k_{4})}{\sin(\pi
k_{1}\cdot k_{2})}A_{tu}^{(p_{n};q_{m};r_{l})}=\frac{\sin(\frac{u}{2}+2-N)\pi
}{\sin(\frac{s}{2}+2-N)\pi}A_{tu}^{(p_{n};q_{m};r_{l})}\nonumber\\
&  =\frac{(-1)^{N}\Gamma(\frac{s}{2}+2-N)\Gamma(\frac{-s}{2}-1+N)}%
{\Gamma(\frac{u}{2}+2)\Gamma(\frac{-u}{2}-1)}\nonumber\\
&  \cdot\int_{1}^{\infty}dx\,x^{k_{1}\cdot k_{2}}(1-x)^{k_{2}\cdot k_{3}}%
\cdot\prod_{n=1}\left[  \frac{\left(  -1\right)  ^{n-1}(n-1)!k_{1}^{T}}{x^{n}%
}+\frac{(-1)^{n-1}(n-1)!k_{3}^{T}}{(x-1)^{n}}\right]  ^{p_{n}}\nonumber\\
&  \cdot\prod_{m=1}\left[  \frac{\left(  -1\right)  ^{m-1}(m-1)!k_{1}^{P}%
}{x^{m}}+\frac{(-1)^{m-1}(m-1)!k_{3}^{P}}{(x-1)^{m}}\right]  ^{q_{m}%
}\nonumber\\
&  \cdot\prod_{l=1}\left[  \frac{\left(  -1\right)  ^{l-1}(l-1)!k_{1}^{L}%
}{x^{l}}+\frac{(-1)^{l-1}(l-1)!k_{3}^{L}}{(x-1)^{l}}\right]  ^{r_{l}%
}\nonumber\\
&  =\frac{(-1)^{N}\Gamma(\frac{s}{2}+2-N)\Gamma(\frac{-s}{2}-1+N)}%
{\Gamma(\frac{u}{2}+2)\Gamma(\frac{-u}{2}-1)}\nonumber\\
&  \int_{1}^{\infty}dx\,x^{k_{1}\cdot k_{2}}(1-x)^{k_{2}\cdot k_{3}-N}%
\cdot\prod_{n=1}\left(  k_{3}^{T}\left(  -1\right)  ^{n-1}(n-1)![1-(\frac
{-k_{1}^{T}}{k_{3}^{T}})(\frac{x-1}{x})^{n}]\right)  ^{p_{n}}\nonumber\\
&  \cdot\prod_{m=1}\left(  k_{3}^{P}\left(  -1\right)  ^{m-1}(m-1)![1-(\frac
{-k_{1}^{P}}{k_{3}^{P}})(\frac{x-1}{x})^{m}]\right)  ^{q_{m}}\nonumber\\
&  \cdot\prod_{l=1}\left(  k_{3}^{L}\left(  -1\right)  ^{l-1}(l-1)![1-(\frac
{-k_{1}^{L}}{k_{3}^{L}})(\frac{x-1}{x})^{l}]\right)  ^{r_{l}}.
\end{align}

\bigskip We can now do a change of variable $\frac{x-1}{x}=y$ to get%

\begin{align}
A_{st}^{(p_{n};q_{m};r_{l})}  &  =\frac{(-1)^{N}\Gamma(\frac{s}{2}%
+2-N)\Gamma(\frac{-s}{2}-1+N)}{\Gamma(\frac{u}{2}+2)\Gamma(\frac{-u}{2}%
-1)}\int_{0}^{1}dy\,y^{k_{2}\cdot k_{3}-N}(1-y)^{-k_{1}\cdot k_{2}-k_{2}\cdot
k_{3}+N-2}\nonumber\\
&  \cdot\prod_{n=1}\left(  k_{3}^{T}\left(  -1\right)  ^{n-1}(n-1)![1-(\frac
{-k_{1}^{T}}{k_{3}^{T}})y^{n}]\right)  ^{p_{n}}\nonumber\\
&  \cdot\prod_{m=1}\left(  k_{3}^{P}\left(  -1\right)  ^{m-1}(m-1)![1-(\frac
{-k_{1}^{P}}{k_{3}^{P}})y^{m}]\right)  ^{q_{m}}\nonumber\\
&  \cdot\prod_{l=1}\left(  k_{3}^{L}\left(  -1\right)  ^{l-1}(l-1)![1-(\frac
{-k_{1}^{L}}{k_{3}^{L}})y^{l}]\right)  ^{r_{l}}\nonumber\\
&  =\frac{(-1)^{N}\Gamma(\frac{s}{2}+2-N)\Gamma(\frac{-s}{2}-1+N)}%
{\Gamma(\frac{u}{2}+2)\Gamma(\frac{-u}{2}-1)}\cdot\prod_{n=1}\left[  \left(
-1\right)  ^{n-1}(n-1)!k_{3}^{T}\right]  ^{p_{n}}\nonumber\\
&  \prod_{m=1}\left[  \left(  -1\right)  ^{m-1}(m-1)!k_{3}^{P}\right]
^{q_{m}}\prod_{l=1}\left[  \left(  -1\right)  ^{l-1}(l-1)!k_{3}^{L}\right]
^{r_{l}}\nonumber\\
&  \cdot\int_{0}^{1}dy\,y^{k_{2}\cdot k_{3}-N}(1-y)^{-k_{1}\cdot k_{2}%
-k_{2}\cdot k_{3}+N-2}\nonumber\\
&  \cdot\left(  1-(z_{n}^{T}y)^{n}\right)  ^{p_{n}}\left(  1-(z_{m}^{P}%
y)^{m}\right)  ^{q_{m}}\left(  1-(z_{l}^{L}y)^{l}\right)  ^{r_{l}}.
\end{align}
Finally the amplitude can be written in the following form%

\begin{align}
A_{st}^{(p_{n};q_{m};r_{l})}  &  =\frac{\Gamma(\frac{s}{2}+2-N)\Gamma
(\frac{-s}{2}-1+N)}{\Gamma(\frac{u}{2}+2)\Gamma(\frac{-u}{2}-1)}\prod
_{n=1}\left[  -(n-1)!k_{3}^{T}\right]  ^{p_{n}}\nonumber\\
&  \cdot\prod_{m=1}\left[  -(m-1)!k_{3}^{P}\right]  ^{q_{m}}\prod_{l=1}\left[
-(l-1)!k_{3}^{L}\right]  ^{r_{l}}\nonumber\\
&  \cdot\int_{0}^{1}dy\,y^{\frac{-t}{2}-2}(1-y)^{\frac{-u}{2}-2}[(1-z_{n}%
^{T}y)(1-z_{n}^{T}\omega_{n}{}^{2}y)...(1-z_{n}^{T}\omega_{n}^{n-1}y)]^{p_{n}%
}\nonumber\\
&  \cdot\lbrack(1-z_{m}^{P}y)(1-z_{m}^{P}\omega_{m}y)...(1-z_{m}^{P}\omega
_{m}^{m-1}y)]^{q_{m}}\nonumber\\
&  \cdot\lbrack(1-z_{l}^{L}y)(1-z_{l}^{L}\omega_{l}y)...(1-w_{l}^{L}\omega
_{l}^{l-1}y)]^{p_{n}},
\end{align}
which can then be written in terms of the D-type Lauricella function
$F_{D}^{(K)}$ as following%
\begin{align}
&  A_{st}^{(p_{n};q_{m};r_{l})}\nonumber\\
&  =\frac{\Gamma(\frac{s}{2}+2-N)\Gamma(\frac{-s}{2}-1+N)}{\Gamma(\frac{u}%
{2}+2)\Gamma(\frac{-u}{2}-1)}\frac{\Gamma(\frac{-t}{2}-1)\Gamma(\frac{-u}%
{2}-1)}{\Gamma(\frac{s}{2}+2-N)}\nonumber\\
&  \cdot\prod_{n=1}\left[  -(n-1)!k_{3}^{T}\right]  ^{p_{n}}\prod_{m=1}\left[
-(m-1)!k_{3}^{P}\right]  ^{q_{m}}\prod_{l=1}\left[  -(l-1)!k_{3}^{L}\right]
^{r_{l}}\nonumber\\
&  \cdot F_{D}^{(K)}\left(
\begin{array}
[c]{c}%
-\frac{t}{2}-1;\left\{  -p_{1}\right\}  ^{1},...,\left\{  -p_{n}\right\}
^{n},\left\{  -q_{1}\right\}  ^{1},...,\left\{  -q_{m}\right\}  ^{m},\left\{
-r_{1}\right\}  ^{1},...,\left\{  -r_{l}\right\}  ^{l};\frac{s}{2}+2-N;\\
\left[  z_{1}^{T}\right]  ,...,\left[  z_{n}^{T}\right]  ,\left[  z_{1}%
^{P}\right]  ,...,\left[  z_{m}^{P}\right]  ,\left[  z_{1}^{L}\right]
,...,\left[  z_{l}^{L}\right]  ,
\end{array}
\right) \nonumber\\
&  =\frac{\Gamma(\frac{-s}{2}-1+N)\Gamma(\frac{-t}{2}-1)}{\Gamma(\frac{u}%
{2}+2)}\prod_{n=1}\left[  -(n-1)!k_{3}^{T}\right]  ^{p_{n}}\prod_{m=1}\left[
-(m-1)!k_{3}^{P}\right]  ^{q_{m}}\prod_{l=1}\left[  -(l-1)!k_{3}^{L}\right]
^{r_{l}}\nonumber\\
&  \cdot F_{D}^{(K)}\left(
\begin{array}
[c]{c}%
-\frac{t}{2}-1;\left\{  -p_{1}\right\}  ^{1},...,\left\{  -p_{n}\right\}
^{n},\left\{  -q_{1}\right\}  ^{1},...,\left\{  -q_{m}\right\}  ^{m},\left\{
-r_{1}\right\}  ^{1},...,\left\{  -r_{l}\right\}  ^{l};\frac{s}{2}+2-N;\\
\left[  z_{1}^{T}\right]  ,...,\left[  z_{n}^{T}\right]  ,\left[  z_{1}%
^{P}\right]  ,...,\left[  z_{m}^{P}\right]  ,\left[  z_{1}^{L}\right]
,...,\left[  z_{l}^{L}\right]
\end{array}
\right)  \label{2222}%
\end{align}
where we have defined%
\begin{equation}
k_{i}^{X}\equiv e^{X}\cdot k_{i}\text{, }\omega_{k}=e^{\frac{2\pi i}{k}%
}\text{\ , \ \ }z_{k}^{X}=(\frac{-k_{1}^{X}}{k_{3}^{X}})^{\frac{1}{k}}%
\end{equation}
and%
\begin{equation}
\left\{  a\right\}  ^{n}=\underset{n}{\underbrace{a,a,\cdots,a}}\text{,
\ \ }\left[  z_{k}^{X}\right]  =z_{k}^{X},z_{k}^{X}e^{\frac{2\pi i}{k}}%
,\cdots,z_{k}^{X}e^{\frac{2\pi i(k-1)}{k}}\text{ or }z_{k}^{X},z_{k}^{X}%
\omega_{k},...,z_{k}^{X}\omega_{k}^{k-1}.
\end{equation}
The integer $K$ in Eq.(\ref{2222}) is defined to be%
\begin{equation}
\text{ }K=\underset{\{\text{for all }r_{j}^{T}\neq0\}}{\sum_{j=1}^{n}%
j}+\underset{\{\text{for all }r_{j}^{P}\neq0\}}{\sum_{j=1}^{m}j}%
+\underset{\{\text{for all }r_{j}^{L}\neq0\}}{\sum_{j=1}^{l}j}.
\end{equation}
For a given $K$, there can be SSA with different mass level $N$.

Alternatively, by using the identity of Lauricella function for $b_{i}\in
Z^{-}$%
\begin{align}
&  F_{D}^{(K)}\left(  a;b_{1},...,b_{K};c;x_{1},...,x_{K}\right)
=\frac{\Gamma\left(  c\right)  \Gamma\left(  c-a-\sum b_{i}\right)  }%
{\Gamma\left(  c-a\right)  \Gamma\left(  c-\sum b_{i}\right)  }\nonumber\\
\cdot &  F_{D}^{(K)}\left(  a;b_{1},...,b_{K};1+a+\sum b_{i}-c;1-x_{1}%
,...,1-x_{K}\right)  ,
\end{align}
we can rederive the string BCJ relation \cite{stringBCJ, stringBCJ2,LLY}%
\begin{align}
\frac{A_{st}^{(r_{n}^{T},r_{m}^{P},r_{l}^{L})}}{A_{tu}^{(r_{n}^{T},r_{m}%
^{P},r_{l}^{L})}} &  =\frac{(-)^{N}\Gamma\left(  -\frac{s}{2}-1\right)
\Gamma\left(  \frac{s}{2}+2\right)  }{\Gamma\left(  \frac{u}{2}+2-N\right)
\Gamma\left(  -\frac{u}{2}-1+N\right)  }\nonumber\\
&  =\frac{\sin\left(  \frac{\pi u}{2}\right)  }{\sin\left(  \frac{\pi s}%
{2}\right)  }=\frac{\sin\left(  \pi k_{2}\cdot k_{4}\right)  }{\sin\left(  \pi
k_{1}\cdot k_{2}\right)  },\label{BCJ}%
\end{align}
which gives another form of the $\left(  s,t\right)  $ channel amplitude%
\begin{align}
&  A_{st}^{(r_{n}^{T},r_{m}^{P},r_{l}^{L})}\nonumber\\
&  =B\left(  -\frac{t}{2}-1,-\frac{s}{2}-1\right)  \prod_{n=1}\left[
-(n-1)!k_{3}^{T}\right]  ^{r_{n}^{T}}\nonumber\\
&  \cdot\prod_{m=1}\left[  -(m-1)!k_{3}^{P}\right]  ^{r_{m}^{P}}\prod
_{l=1}\left[  -(l-1)!k_{3}^{L}\right]  ^{r_{l}^{L}}\nonumber\\
&  \cdot F_{D}^{(K)}\left(  -\frac{t}{2}-1;R_{n}^{T},R_{m}^{P},R_{l}^{L}%
;\frac{u}{2}+2-N;\tilde{Z}_{n}^{T},\tilde{Z}_{m}^{P},\tilde{Z}_{l}^{L}\right)
\label{st}%
\end{align}
and similarly the $\left(  t,u\right)  $ channel amplitude
\begin{align}
&  A_{tu}^{(r_{n}^{T},r_{m}^{P},r_{l}^{L})}\nonumber\\
&  =B\left(  -\frac{t}{2}-1,-\frac{u}{2}-1\right)  \prod_{n=1}\left[
-(n-1)!k_{3}^{T}\right]  ^{r_{n}^{T}}\nonumber\\
&  \cdot\prod_{m=1}\left[  -(m-1)!k_{3}^{P}\right]  ^{r_{m}^{P}}\prod
_{l=1}\left[  -(l-1)!k_{3}^{L}\right]  ^{r_{l}^{L}}\nonumber\\
&  \cdot F_{D}^{(K)}\left(  -\frac{t}{2}-1;R_{n}^{T},R_{m}^{P},R_{l}^{L}%
;\frac{s}{2}+2-N;Z_{n}^{T},Z_{m}^{P},Z_{l}^{L}\right)  .\label{tu2}%
\end{align}
In Eq.(\ref{st}) and Eq.(\ref{tu2}), we have defined%
\begin{equation}
R_{k}^{X}\equiv\left\{  -r_{1}^{X}\right\}  ^{1},\cdots,\left\{  -r_{k}%
^{X}\right\}  ^{k}\text{ with }\left\{  a\right\}  ^{n}%
=\underset{n}{\underbrace{a,a,\cdots,a}},
\end{equation}
and%
\begin{equation}
Z_{k}^{X}\equiv\left[  z_{1}^{X}\right]  ,\cdots,\left[  z_{k}^{X}\right]
\text{ with }\left[  z_{k}^{X}\right]  =z_{k0}^{X},\cdots,z_{k\left(
k-1\right)  }^{X}%
\end{equation}
where%
\begin{equation}
z_{k}^{X}=\left\vert \left(  -\frac{k_{1}^{X}}{k_{3}^{X}}\right)  ^{\frac
{1}{k}}\right\vert ,\ z_{kk^{\prime}}^{X}=z_{k}^{X}e^{\frac{2\pi ik^{\prime}%
}{k}},\ \tilde{z}_{kk^{\prime}}^{X}\equiv1-z_{kk^{\prime}}^{X}%
\end{equation}
for $k^{\prime}=0,\cdots,k-1.$

With the notation introduced above, the $(s,t)$ channel amplitude in
Eq.(\ref{2222}) can be rewritten as%
\begin{align}
&  A_{st}^{(r_{n}^{T},r_{m}^{P},r_{l}^{L})}\nonumber\\
&  =B\left(  -\frac{t}{2}-1,-\frac{s}{2}-1+N\right)  \prod_{n=1}\left[
-(n-1)!k_{3}^{T}\right]  ^{r_{n}^{T}}\nonumber\\
&  \cdot\prod_{m=1}\left[  -(m-1)!k_{3}^{P}\right]  ^{r_{m}^{P}}\prod
_{l=1}\left[  -(l-1)!k_{3}^{L}\right]  ^{r_{l}^{L}}\nonumber\\
&  \cdot F_{D}^{(K)}\left(  -\frac{t}{2}-1;R_{n}^{T},R_{m}^{P},R_{l}^{L}%
;\frac{s}{2}+2-N;Z_{n}^{T},Z_{m}^{P},Z_{l}^{L}\right)  . \label{tu}%
\end{align}

\section{Regge scattering limit}

With the exact SSA calculated in Eq.(\ref{tu}), Eq.(\ref{st}) and
Eq.(\ref{tu2}) which are valid for all kinematic regimes, we can rederive SSA
and symmetries or relations among SSA of different string states at various
limits calculated previously. These include the linear relations conjectured
by Gross \cite{GM,Gross,GrossManes} and proved in
\cite{ChanLee1,ChanLee2,CHL,PRL, CHLTY,susy} in the hard scattering limit, the
recurrence relations in the Regge scattering limit \cite{KLY,LY,LY2014} and
the extended recurrence relations in the nonrelativistic scattering limit
\cite{LLY} discovered recently. In this section, we first calculate the Regge
scattering limit. The relevant kinematics in Regge limit are%
\begin{align}
k_{1}^{T}  &  =0\text{, \ \ }k_{3}^{T}\simeq-\sqrt{-t},\\
k_{1}^{P}  &  \simeq-\frac{s}{2M_{2}}\text{,\ }k_{3}^{P}\simeq-\frac{\tilde
{t}}{2M_{2}}=-\frac{t-M_{2}^{2}-M_{3}^{2}}{2M_{2}},\\
k_{1}^{L}  &  \simeq-\frac{s}{2M_{2}}\text{, }k_{3}^{L}\simeq-\frac{\tilde
{t}^{\prime}}{2M_{2}}=-\frac{t+M_{2}^{2}-M_{3}^{2}}{2M_{2}}.
\end{align}
One can easily calculate%
\begin{equation}
\tilde{z}_{kk^{\prime}}^{T}=1,\ \tilde{z}_{kk^{\prime}}^{P}=1-\left(
-\frac{s}{\tilde{t}}\right)  ^{1/k}e^{\frac{i2\pi k^{\prime}}{k}}\sim s^{1/k}%
\end{equation}
and%
\begin{equation}
\tilde{z}_{kk^{\prime}}^{L}=1-\left(  -\frac{s}{\tilde{t}^{\prime}}\right)
^{1/k}e^{\frac{i2\pi k^{\prime}}{k}}\sim s^{1/k}.
\end{equation}
In the Regge limit, the SSA in Eq.(\ref{st}) reduces to%
\begin{align}
&  A_{st}^{(r_{n}^{T},r_{m}^{P},r_{l}^{L})}\nonumber\\
\simeq &  B\left(  -\frac{t}{2}-1,-\frac{s}{2}-1\right)  \prod_{n=1}\left[
(n-1)!\sqrt{-t}\right]  ^{r_{n}^{T}}\nonumber\\
\cdot &  \prod_{m=1}\left[  (m-1)!\frac{\tilde{t}}{2M_{2}}\right]  ^{r_{m}%
^{P}}\prod_{l=1}\left[  (l-1)!\frac{\tilde{t}^{\prime}}{2M_{2}}\right]
^{r_{l}^{L}}\nonumber\\
\cdot &  F_{1}\left(  -\frac{t}{2}-1;-q_{1},-r_{1};-\frac{s}{2};\frac
{s}{\tilde{t}},\frac{s}{\tilde{t}^{\prime}}\right)  . \label{app}%
\end{align}
where $F_{1}$ is the Appell function. Eq.(\ref{app}) agrees with the result
obtained in \cite{LY2014} previously.

\section{Hard scattering limit}

In this section, we rederive the linear relations conjectured by Gross
\cite{GM,Gross,GrossManes} and corrected and proved in
\cite{ChanLee1,ChanLee2,CHL,PRL, CHLTY,susy} in the hard scattering limit. As
we will see that the calculation will be more subtle than that of the Regge
scattering limit. In the hard scattering limit $e^{P}=e^{L}$
\cite{ChanLee1,ChanLee2}, and we can consider only the polarization $e^{L}$
case. We first briefly review the results \cite{review} for linear relations
among hard SSA. One first observes that for each fixed mass level $N$ only
states of the following form \cite{PRL,CHLTY}
\begin{equation}
\left\vert N,2m,q\right\rangle \equiv(\alpha_{-1}^{T})^{N-2m-2q}(\alpha
_{-1}^{L})^{2m}(\alpha_{-2}^{L})^{q}|0,k\rangle\label{Nmq}%
\end{equation}
are of leading order in energy in the HSS limit. The choice of only even power
$2m$ in $\alpha_{-1}^{L}$ is the result of the observation
\cite{ChanLee1,ChanLee2} that the naive energy order of the amplitudes will in
general drop by even number of energy powers. Scattering amplitudes
corresponding to states with $(\alpha_{-1}^{L})^{2m+1}$\ turn out to be of
subleading order in energy. Many simplifications occur if we apply Ward
identities or decoupling of ZNS only on high energy states in Eq.(\ref{Nmq})
in the HSS limit. One important result was the discovery of the linear
relations among hard SSA of different string states at each fixed mass level
$N$ \cite{PRL,CHLTY}
\begin{equation}
\frac{A_{st}^{(N,2m,q)}}{A_{st}^{(N,0,0)}}=\left(  -\frac{1}{M_{2}}\right)
^{2m+q}\left(  \frac{1}{2}\right)  ^{m+q}(2m-1)!!. \label{04}%
\end{equation}
Exactly the same results can also be obtained by two other calculations, the
Virasoro constraint calculation and the corrected saddle-point calculation
\cite{PRL,CHLTY}. In the decoupling of ZNS calculations at the mass level
$M_{2}^{2}=4$, for example, there are four leading order SSA
\cite{ChanLee1,ChanLee2}
\begin{equation}
A_{TTT}:A_{LLT}:A_{(LT)}:A_{[LT]}=8:1:-1:-1 \label{03}%
\end{equation}
which are proportional to each other. While the saddle point calculation of
\cite{GrossManes} gave $A_{TTT}\propto A_{[LT]},$ and $A_{LLT}=0$ which are
inconsistent with the decoupling of ZNS or unitarity of the theory. Indeed, a
sample calculation was done \cite{ChanLee1,ChanLee2} to explicitly verify the
ratios in Eq.(\ref{03}).

One interesting application of Eq.(\ref{04}) was the derivation of relation of
$A_{st}^{(N,2m,q)}$ and $A_{tu}^{(N,2m,q)}$ in the hard scattering limit
\cite{Closed}%
\begin{equation}
A_{st}^{(N,2m,q)}\simeq(-)^{N}\frac{\sin(\pi k_{2}\cdot k_{4})}{\sin(\pi
k_{1}\cdot k_{2})}A_{tu}^{(N,2m,q)}\label{HBCJ}%
\end{equation}
where
\begin{align}
A_{tu}^{(N,2m,q)} &  \simeq\sqrt{\pi}(-1)^{N-1}2^{-N}E^{-1-2N}\left(
\sin\frac{\phi}{2}\right)  ^{-3}\left(  \cos\frac{\phi}{2}\right)
^{5-2N}\nonumber\\
&  \cdot\exp\left[  -\frac{t\ln t+u\ln u-(t+u)\ln(t+u)}{2}\right]  .
\end{align}
Eq.(\ref{HBCJ}) was shown to be valid for scatterings of four arbitrary string
states and was obtained in 2006 \cite{2006}, and thus was earlier than the
discovery of four point field theory BCJ relations \cite{BCJ} and "string BCJ
relations" in Eq.(\ref{BCJ}) \cite{LLY,stringBCJ, stringBCJ2}. In contrast to
the calculation of string BCJ relations \cite{stringBCJ, stringBCJ2} which was
motivated by the field theory BCJ relations \cite{BCJ}, the derivation of
Eq.(\ref{HBCJ}) was motivated by the calculation of hard closed SSA
\cite{Closed} by using KLT relation \cite{KLT}. See a more detailed discussion
in a recent publication \cite{LLY}.

We are now ready to rederive Eq.(\ref{Nmq}) and Eq.(\ref{04}) from
Eq.(\ref{st}). The relevant kinematics are%
\begin{align}
k_{1}^{T}  &  =0\text{, \ \ }k_{3}^{T}\simeq-E\sin\phi,\\
k_{1}^{L}  &  \simeq-\frac{2p^{2}}{M_{2}}\simeq-\frac{2E^{2}}{M_{2}},\\
k_{3}^{L}  &  \simeq\frac{2E^{2}}{M_{2}}\sin^{2}\frac{\phi}{2}.
\end{align}
One can calculate%
\begin{equation}
\tilde{z}_{kk^{\prime}}^{T}=1,\ \tilde{z}_{kk^{\prime}}^{L}=1-\left(
-\frac{s}{t}\right)  ^{1/k}e^{\frac{i2\pi k^{\prime}}{k}}\sim O\left(
1\right)  .
\end{equation}
The SSA in Eq.(\ref{st}) reduces to%
\begin{align}
&  A_{st}^{(r_{n}^{T},r_{l}^{L})}=B\left(  -\frac{t}{2}-1,-\frac{s}%
{2}-1\right) \nonumber\\
&  \cdot\prod_{n=1}\left[  (n-1)!E\sin\phi\right]  ^{r_{n}^{T}}\prod
_{l=1}\left[  -(l-1)!\frac{2E^{2}}{M_{2}}\sin^{2}\frac{\phi}{2}\right]
^{r_{l}^{L}}\nonumber\\
&  \cdot F_{D}^{(K)}\left(  -\frac{t}{2}-1;R_{n}^{T},R_{l}^{L};\frac{u}%
{2}+2-N;\left(  1\right)  _{n},\tilde{Z}_{l}^{L}\right)  .
\end{align}
As was mentioned above that, in the hard scattering limit, there was a
difference between the naive energy order and the real energy order
corresponding to the $\left(  \alpha_{-1}^{L}\right)  ^{r_{1}^{L}}$ operator
in Eq.(\ref{state}). So let's pay attention to the corresponding summation and
write%
\begin{align}
&  A_{st}^{(r_{n}^{T},r_{l}^{L})}=B\left(  -\frac{t}{2}-1,-\frac{s}%
{2}-1\right) \nonumber\\
&  \cdot\prod_{n=1}\left[  (n-1)!E\sin\phi\right]  ^{r_{n}^{T}}\prod
_{l=1}\left[  -(l-1)!\frac{2E^{2}}{M_{2}}\sin^{2}\frac{\phi}{2}\right]
^{r_{l}^{L}}\nonumber\\
&  \cdot\sum_{k_{r}}\frac{\left(  -\frac{t}{2}-1\right)  _{k_{r}}}{\left(
\frac{u}{2}+2-N\right)  _{k_{r}}}\frac{\left(  -r_{1}^{L}\right)  _{k_{r}}%
}{k_{r}!}\left(  1+\frac{s}{t}\right)  ^{k_{r}}\cdot\left(  \cdots\right)
\end{align}
where we have used $\left(  a\right)  _{n+m}=\left(  a\right)  _{n}\left(
a+n\right)  _{m}$ and $\left(  \cdots\right)  $ are terms which are not
relevant to the following discussion. We then propose the following formula%
\begin{align}
&  \sum_{k_{r}=0}^{r_{1}^{L}}\frac{\left(  -\frac{t}{2}-1\right)  _{k_{r}}%
}{\left(  \frac{u}{2}+2-N\right)  _{k_{r}}}\frac{\left(  -r_{1}^{L}\right)
_{k_{r}}}{k_{r}!}\left(  1+\frac{s}{t}\right)  ^{k_{r}}\nonumber\\
=  &  0\cdot\left(  \frac{tu}{s}\right)  ^{0}\!+0\cdot\left(  \frac{tu}%
{s}\right)  ^{-1}\!+\dots+0\cdot\left(  \frac{tu}{s}\right)  ^{-\left[
\frac{r_{1}^{L}+1}{2}\right]  -1}\nonumber\\
&  +C_{r_{1}^{L}}\left(  \frac{tu}{s}\right)  ^{-\left[  \frac{r_{1}^{L}+1}%
{2}\right]  }+\mathit{O}\left\{  \left(  \frac{tu}{s}\right)  ^{-\left[
\frac{r_{1}^{L}+1}{2}\right]  +1}\right\}  . \label{pro}%
\end{align}
where $C_{r_{1}^{L}}$ is independent of energy $E$ and depends on $r_{1}^{L}$
and possibly scattering angle $\phi$. For $r_{1}^{L}=2m$\ being an even
number, we further propose that $C_{r_{1}^{L}}=\frac{\left(  2m\right)  !}%
{m!}$ and is $\phi$ independent. We have verified Eq.(\ref{pro}) for
$r_{1}^{L}=0,1,2,\cdots,10$.

It should be noted that, taking Regge limit ($s\rightarrow\infty$ with $t$
fixed) and setting $r_{1}^{L}=2m$, Eq.(\ref{pro}) reduces to the Stirling
number identity,%
\begin{gather}
\sum_{k_{r}=0}^{2m}\frac{\left(  -\frac{t}{2}-1\right)  _{k_{r}}}{\left(
-\frac{s}{2}\right)  _{k_{r}}}\frac{\left(  -2m\right)  _{k_{r}}}{k_{r}%
!}\left(  \frac{s}{t}\right)  ^{k_{r}}\simeq\sum_{k_{r}=0}^{2m}\left(
-2m\right)  _{k_{r}}\left(  -\frac{t}{2}-1\right)  _{k_{r}}\frac{\left(
-2/t\right)  ^{k_{r}}}{k_{r}!}\nonumber\\
=0\cdot\left(  -t\right)  ^{0}\!+0\cdot\left(  -t\right)  ^{-1}\!+\dots
+0\cdot\left(  -t\right)  ^{-m+1}+\frac{(2m)!}{m!}\left(  -t\right)
^{-m}+\mathit{O}\left\{  \left(  \frac{1}{t}\right)  ^{m+1}\right\}  ,
\label{stirling}%
\end{gather}
which was proposed in \cite{KLY} and proved in \cite{LYY}.

It was demonstrated in \cite{KLY} that the ratios in the hard scattering limit
in Eq.(\ref{04}) can be reproduced from a class of Regge string scattering
amplitudes presented in Eq.(\ref{app}). The key of the mathematical proof
\cite{LYY} was the new Stirling number identity proposed in Eq.(\ref{stirling}).

In Eq.(\ref{pro}), the $0$ terms correspond to the naive leading energy orders
in the hard SSA calculation. The true leading order SSA in the hard scattering
limit can then be identified%
\begin{align}
&  A_{st}^{(r_{n}^{T},r_{l}^{L})}\simeq B\left(  -\frac{t}{2}-1,-\frac{s}%
{2}-1\right) \nonumber\\
&  \cdot\prod_{n=1}\left[  (n-1)!E\sin\phi\right]  ^{r_{n}^{T}}\prod
_{l=1}\left[  -(l-1)!\frac{2E^{2}}{M_{2}}\sin^{2}\frac{\phi}{2}\right]
^{r_{l}^{L}}\nonumber\\
&  \cdot C_{r_{1}^{L}}\left(  E\sin\phi\right)  ^{-2\left[  \frac{r_{1}^{L}%
+1}{2}\right]  }\cdot\left(  \cdots\right) \nonumber\\
&  \sim E^{N-\sum_{n\geq2}nr_{n}^{T}-\left(  2\left[  \frac{r_{1}^{L}+1}%
{2}\right]  -r_{1}^{L}\right)  -\sum_{l\geq3}lr_{l}^{L}},
\end{align}
which means that SSA reaches its highest energy when $r_{n\geq2}^{T}%
=r_{l\geq3}^{L}=0$ and $r_{1}^{L}=2m$ being an even number. This is consistent
with the previous result presented in Eq.(\ref{Nmq})
\cite{ChanLee1,ChanLee2,CHL,PRL, CHLTY,susy}.

Finally, the leading order SSA in the hard scattering limit, i.e. $r_{1}%
^{T}=N-2m-2$, $r_{1}^{L}=2m$ and $r_{2}^{L}=q$, can be calculated to be%
\begin{align}
&  A_{st}^{(N-2m-2q,2m,q)}\nonumber\\
&  \simeq B\left(  -\frac{t}{2}-1,-\frac{s}{2}-1\right)  \left(  E\sin
\phi\right)  ^{N}\frac{\left(  2m\right)  !}{m!}\left(  -\frac{1}{2M_{2}%
}\right)  ^{2m+q}\nonumber\\
&  =(2m-1)!!\left(  -\frac{1}{M_{2}}\right)  ^{2m+q}\left(  \frac{1}%
{2}\right)  ^{m+q}A_{st}^{(N,0,0)}%
\end{align}
which reproduces the ratios in Eq.(\ref{04}), and is consistent with the
previous result \cite{ChanLee1,ChanLee2,CHL,PRL, CHLTY,susy}.

\section{Nonrelativistic scattering limit}

In a recent paper \cite{LLY} both $s-t$ and $t-u$ channel nonrelativistic low
energy string scattering amplitudes of three tachyons and one leading
trajectory string state at arbitrary mass levels were calculated. It was
discovered that the mass and spin dependent nonrelativistic string BCJ
relations \cite{stringBCJ, stringBCJ2} can be expressed in terms of Gauss
hypergeometric functions. As an application, for each fixed mass level $N,$
the extended recurrence relations among nonrelativistic low energy string
scattering amplitudes of string states with different spins and different
channels can be derived.

In this section, we intend to rederive the results stated above from the
Lauricella functions. In the nonrelativistic limit $|\vec{k_{1}}|\ll M_{2}$,
we have%
\begin{align}
k_{1}^{T}  &  =0,k_{3}^{T}=-\left[  \frac{\epsilon}{2}+\frac{(M_{1}+M_{2}%
)^{2}}{4M_{1}M_{2}\epsilon}|\vec{k_{1}}|^{2}\right]  \sin\phi,\\
k_{1}^{L}  &  =-\frac{M_{1}+M_{2}}{M_{2}}|\vec{k_{1}}|+O\left(  |\vec{k_{1}%
}|^{2}\right)  ,\\
k_{3}^{L}  &  =-\frac{\epsilon}{2}\cos\phi+\frac{M_{1}+M_{2}}{2M_{2}}%
|\vec{k_{1}}|+O\left(  |\vec{k_{1}}|^{2}\right)  ,\\
k_{1}^{P}  &  =-M_{1}+O\left(  |\vec{k_{1}}|^{2}\right)  ,\\
k_{3}^{P}  &  =\frac{M_{1}+M_{2}}{2}-\frac{\epsilon}{2M_{2}}\cos\phi
|\vec{k_{1}}|+O\left(  |\vec{k_{1}}|^{2}\right)
\end{align}
where $\epsilon=\sqrt{(M_{1}+M_{2})^{2}-4M_{3}^{2}}$. One can easily calculate%
\begin{equation}
z_{k}^{T}=z_{k}^{L}=0,z_{k}^{P}\simeq\left\vert \left(  \frac{2M_{1}}%
{M_{1}+M_{2}}\right)  ^{\frac{1}{k}}\right\vert .
\end{equation}
The SSA in Eq.(\ref{tu}) reduces to%
\begin{align}
&  A_{st}^{(r_{n}^{T},r_{m}^{P},r_{l}^{L})}\nonumber\\
&  \simeq\prod_{n=1}\left[  (n-1)!\frac{\epsilon}{2}\sin\phi\right]
^{r_{n}^{T}}\prod_{m=1}\left[  -(m-1)!\frac{M_{1}+M_{2}}{2}\right]
^{r_{m}^{P}}\nonumber\\
&  \cdot\prod_{l=1}\left[  (l-1)!\frac{\epsilon}{2}\cos\phi\right]
^{r_{l}^{L}}B\left(  \frac{M_{1}M_{2}}{2},1-M_{1}M_{2}\right) \nonumber\\
&  \cdot F_{D}^{(K)}\left(  \frac{M_{1}M_{2}}{2};R_{m}^{P};M_{1}M_{2};\left(
\frac{2M_{1}}{M_{1}+M_{2}}\right)  _{m}\right)
\end{align}
where%
\begin{equation}
K=\underset{\{\text{for all }r_{j}^{P}\neq0\}}{\sum_{j=1}^{m}j}.
\end{equation}

Note that for string states with $r_{k}^{P}=0$ for all $k\geq2$, one has $K=1$
and the Lauricella functions in the low energy nonrelativistic SSA reduce to
the Gauss hypergeometric functions $F_{D}^{(1)}=$ $_{2}F_{1}.$ In particular,
for the case of $r_{1}^{T}=N_{1}$, $r_{1}^{P}=N_{3}$, $r_{1}^{L}=N_{2}$, and
$r_{k}^{X}=0$ for all $k\geq2$, the SSA reduces to%
\begin{align}
&  A_{st}^{(N_{1},N_{2},N_{3})}=\left(  \frac{\epsilon}{2}\sin\phi\right)
^{N_{1}}\left(  \frac{\epsilon}{2}\cos\phi\right)  ^{N_{2}}\nonumber\\
\cdot &  \left(  -\frac{M_{1}+M_{2}}{2}\right)  ^{N_{3}}B\left(  \frac
{M_{1}M_{2}}{2},1-M_{1}M_{2}\right) \nonumber\\
\cdot &  _{2}F_{1}\left(  \frac{M_{1}M_{2}}{2};-N_{3};M_{1}M_{2};\frac{2M_{1}%
}{M_{1}+M_{2}}\right)  , \label{low}%
\end{align}
which agrees with the result obtained in \cite{LLY} previously. Similarly, one
can calculate the corresponding nonrelativistic $t-u$ channel amplitude as%
\begin{align}
A_{tu}^{(N_{1},N_{2},N_{3})}=  &  \left(  -1\right)  ^{N}\left(
\frac{\epsilon}{2}\sin\phi\right)  ^{N_{1}}\left(  \frac{\epsilon}{2}\cos
\phi\right)  ^{N_{2}}\nonumber\\
&  \cdot\left(  -\frac{M_{1}+M_{2}}{2}\right)  ^{N_{3}}B\left(  \frac
{M_{1}M_{2}}{2},\frac{M_{1}M_{2}}{2}\right) \nonumber\\
&  \cdot\text{ }_{2}F_{1}\left(  \frac{M_{1}M_{2}}{2};-N_{3};M_{1}M_{2}%
;\frac{2M_{1}}{M_{1}+M_{2}}\right)  .
\end{align}
Finally the ratio of $s-t$ and $t-u$ channel amplitudes is \cite{LLY}%

\begin{align}
\frac{A_{st}^{(p,r,q)}}{A_{tu}^{(p,r,q)}}  &  =\left(  -1\right)  ^{N}%
\frac{B\left(  -M_{1}M_{2}+1,\frac{M_{1}M_{2}}{2}\right)  }{B\left(
\frac{M_{1}M_{2}}{2},\frac{M_{1}M_{2}}{2}\right)  }\nonumber\\
&  =(-1)^{N}\frac{\Gamma\left(  M_{1}M_{2}\right)  \Gamma\left(  -M_{1}%
M_{2}+1\right)  }{\Gamma\left(  \frac{M_{1}M_{2}}{2}\right)  \Gamma\left(
-\frac{M_{1}M_{2}}{2}+1\right)  }\simeq\frac{\sin\pi\left(  k_{2}\cdot
k_{4}\right)  }{\sin\pi\left(  k_{1}\cdot k_{2}\right)  } \label{NBCJ}%
\end{align}
where, in the nonrelativistic limit, we have%
\begin{subequations}
\begin{align}
k_{1}\cdot k_{2}  &  \simeq-M_{1}M_{2},\\
k_{2}\cdot k_{4}  &  \simeq\frac{\left(  M_{1}+M_{2}\right)  M_{2}}{2}.
\end{align}

We thus have ended up with a consistent nonrelativistic string BCJ relations.
We stress that the above relation is the stringy generalization of the
massless field theory BCJ relation \cite{BCJ} to the higher spin stringy particles.

\section{\noindent The associate symmetry group of string scattering
amplitudes}

In the Lie group approach of special functions, the associate Lie group for
the Lauricella function $F_{D}^{(K)}$ in the SSA at each fixed $K$ is the
$SL\left(  K+3,%
\mathbb{C}
\right)  $\ group \cite{sl4c} which contains the $SL\left(  2,%
\mathbb{C}
\right)  $\ fundamental representation of the $3+1$ dimensional spacetime
Lorentz group $SO(3,1)$. So $sl\left(  K+3,%
\mathbb{C}
\right)  $ contains the $2+1$ dimensional $so(2,1)$ Lorentz spacetime symmetry
on the scattering plane in our case as well. In the Regge limit, the
Lauricella function in the SSA reduces to the Appell function $F_{1}$ with
associate group $SL\left(  5,%
\mathbb{C}
\right)  $ \cite{sl5c}, which is $K$ independent. In the low energy
nonrelativistic limit, the Lauricella function in the SSA reduces to the Gauss
hypergeometric function $_{2}F_{1}$ with associate group $SL\left(  4,%
\mathbb{C}
\right)  $ \cite{sl4c}, which is also $K$ independent.

In sum, we have identified the associate exact $SL\left(  K+3,%
\mathbb{C}
\right)  $ symmetry of string scattering amplitudes with three tachyons and
one\textit{ arbitrary} string states of $26D$ bosonic open string theory.
However, since \textit{not} all Lauricella functions $F_{D}^{(K)}$ with
arbitrary \textit{independent} arguments can be used to represent SSA, it
remained to be studied how the basis states of each $SL(K+3,%
\mathbb{C}
)$ group representation for a given $K$ relates to SSA. This important issue
is currently under investigation.

Finally, with the $SL\left(  K+3,%
\mathbb{C}
\right)  $\ group and the recurrence relations of the Lauricella functions
$F_{D}^{(K)}$, one can derive infinite number of recurrence relations of SSA
of different string states which are valid for \textit{all} energies, as long
as all the Lauricella functions $F_{D}^{(K)}$ in the recurrence relation
representing the SSA. For a simple example, the following recurrence relation
of $F_{D}^{(K)}$ can be verified%
\end{subequations}
\begin{align}
cF_{D}^{(K)}\left(  b_{j};c\right)  +c(x_{j}-1)F_{D}^{(K)}\left(
b_{j}+1;c\right)   & \nonumber\\
+(a-c)x_{j}F_{D}^{(K)}\left(  b_{j}+1;c+1\right)   &  =0, \label{fd}%
\end{align}
which leads to the recurrence relation of SSA%
\begin{equation}
\left(  \frac{u}{2}+2-N\right)  A_{st}^{(r_{n}^{T},r_{m}^{P},r_{l}^{L}%
)}-\left(  \frac{s}{2}+1\right)  k_{3}^{T}A_{st}^{(r_{n}^{\prime T},r_{m}%
^{P},r_{l}^{L})}=0
\end{equation}
where $(r_{n}^{\prime T},r_{m}^{P},r_{l}^{L})$ means the group $\left(
-\{r_{1}^{T}-1\}^{1},\left\{  -r_{2}^{T}\right\}  ^{2},\cdots,\left\{
-r_{n}^{T}\right\}  ^{n};R_{m}^{P},R_{l}^{L}\right)  $ of polarizations. In
Eq.(\ref{fd}), we have omitted those arguments of $F_{D}^{(K)}$ which remain
the same for all three Lauricella functions.

\begin{acknowledgments}
J.C. would like to thank H. Kawai for crucial suggestions of some results of
this work. This work is supported in part by the Ministry of Science and
Technology and S.T. Yau center of NCTU, Taiwan.
\end{acknowledgments}


\begin{thebibliography}{99}
\bibitem {2006}There was an $(-)^{N}$ ambiguity in Eq.(42) in the high energy
calculation of $(s;t)$ channel amplitude in Ref [31] due to the pole structure
of the gamma function $\Gamma(x)$ for negative real $x.$
\end{thebibliography}
\end{document}